\documentclass[11pt]{article}

\usepackage[letterpaper, textwidth=6in, top=1.1in, bottom=1.15in,
            headsep=0.3in, footskip=0.45in]{geometry}

\usepackage[T1]{fontenc}
\usepackage[utf8]{inputenc}
\usepackage{amsmath}
\usepackage{amsthm}
\usepackage{amssymb}                    
\usepackage{libertinus}                  
\usepackage[semibold, scale=0.94]{sourcesanspro}

\usepackage[varqu, varl, scaled=0.95]{inconsolata}
\usepackage[final]{microtype}

\usepackage[round]{natbib}
\usepackage{url}
\usepackage{booktabs}
\usepackage{nicefrac}
\usepackage{xcolor}
\usepackage{graphicx}
\usepackage{wrapfig}
\usepackage{tcolorbox}
\tcbuselibrary{skins}
\usepackage{tabularx}
\usepackage{tikz}
\definecolor{themeCyan}{HTML}{21918C}
\definecolor{themeIndigo}{HTML}{4B0082} 
\usetikzlibrary{arrows.meta,calc,positioning,decorations.pathreplacing}
\colorlet{internalc}{themeIndigo}
\colorlet{externalc}{themeCyan}
\definecolor{leverc}{HTML}{C45500}
\DeclareRobustCommand{\lever}[1]{\tikz[baseline=(c.base)]\node[circle,draw=leverc,fill=leverc!12,text=leverc,inner sep=0.6pt,font=\scriptsize\bfseries] (c) {#1};}

\makeatletter
\tcbset{height/.code={\tcb@fixedheighttrue
  \tcb@dim@to\kvtcb@height@fixed{#1+0.55cm}\tcb@@ch@@operation{\tcb@ch@fixed}}}
\makeatother
\renewenvironment{wraptable}[2]
  {\begin{table}[t]\centering\captionsetup{width=0.62\linewidth}}
  {\end{table}}

\definecolor{themeCyan}{HTML}{21918C}
\definecolor{themeIndigo}{HTML}{4B0082}
\definecolor{inkDark}{HTML}{1C1A24}
\definecolor{inkMuted}{HTML}{6B6878}
\definecolor{ruleGray}{HTML}{D9D6E0}
\definecolor{abstractBg}{HTML}{F6F3FA}
\definecolor{linkBlue}{HTML}{1D6F8C}

\usepackage{enumitem}
\setlist{itemsep=0.15em, topsep=0.3em}

\usepackage{titlesec}
\titleformat{\section}
  {\sffamily\Large\bfseries\color{themeIndigo}}
  {\color{themeCyan}\thesection}{0.7em}{}
\titleformat{\subsection}
  {\sffamily\large\bfseries\color{themeIndigo}}
  {\color{themeCyan}\thesubsection}{0.65em}{}
\titleformat{\subsubsection}
  {\sffamily\normalsize\bfseries\color{inkDark}}
  {\thesubsubsection}{0.6em}{}
\titlespacing*{\section}{0pt}{1.5em plus 0.4em minus 0.2em}{0.5em}
\titlespacing*{\subsection}{0pt}{1.15em plus 0.3em minus 0.2em}{0.35em}
\titlespacing*{\subsubsection}{0pt}{1em plus 0.2em}{0.25em}

\usepackage{caption}
\usepackage{fancyhdr}
\fancypagestyle{plain}{%
  \fancyhf{}%
  \fancyfoot[C]{\sffamily\small\color{inkMuted}\thepage}%
  }

\renewcommand{\footnoterule}{{\color{ruleGray}%
  \kern-3pt\hrule width 0.3\columnwidth height 0.4pt\kern 2.6pt}}

\usepackage{hyperref}
\hypersetup{
  colorlinks=true,
  linkcolor=themeIndigo,
  citecolor=linkBlue,
  urlcolor=linkBlue,
  pdftitle={Watermarking Should Be Treated as a Monitoring Primitive},
  pdfauthor={Toluwani Aremu, Jie Zhang, Nils Lukas},
  pdfsubject={Position paper, NeurIPS 2026},
}
\usepackage{cleveref}

\newcommand{\venuenote}{Accepted to the Position Paper Track of the
  40th Conference on Neural Information Processing Systems (NeurIPS 2026).}

\newcommand{\makepreprinttitle}{%
  \thispagestyle{plain}%
  \begingroup
  \raggedright
  {\sffamily\small\bfseries\color{themeCyan}%
    \textls[90]{\MakeUppercase{Position Paper}}%
    \hspace{0.7em}{\color{ruleGray}|}\hspace{0.7em}%
    \textls[90]{NEURIPS 2026}\par}
  \vspace{0.9em}
  {\sffamily\fontsize{24}{29}\selectfont\bfseries\color{themeIndigo}%
    Watermarking Should Be Treated as a\\ Monitoring Primitive\par}
  \vspace{1.1em}
  {\color{themeCyan}\rule{2.5cm}{1.6pt}\par}
  \vspace{1.0em}
  {\large\color{inkDark}%
    Toluwani Aremu\textsuperscript{\,1,2}\hspace{1.4em}%
    Jie Zhang\textsuperscript{\,2}\hspace{1.4em}%
    Nils Lukas\textsuperscript{\,1}\par}
  \vspace{0.35em}
  {\small\color{inkMuted}%
    \textsuperscript{1}MBZUAI, UAE\hspace{1.2em}%
    \textsuperscript{2}A*STAR, Singapore\par}
  \endgroup
  \vspace{1.6em}%
  \begingroup
    \let\thefootnote\relax
    \footnotetext{\hspace*{-1.8em}\sffamily\color{inkMuted}\venuenote}%
  \endgroup
}

\newenvironment{preprintabstract}{%
  \begin{tcolorbox}[
    enhanced,
    colback=abstractBg, colframe=abstractBg,
    borderline west={2.2pt}{0pt}{themeIndigo},
    sharp corners, boxrule=0pt,
    left=5mm, right=5mm, top=3.2mm, bottom=3.4mm,
    before skip=0pt, after skip=1.4em]
  {\sffamily\small\bfseries\color{themeIndigo}%
    \MakeUppercase{Abstract}\par}
  \vspace{0.35em}
  \small\setlength{\parskip}{0.4em}%
}{%
  \end{tcolorbox}%
}

\begin{document}

\makepreprinttitle

\begin{preprintabstract}
Watermarking is a primary means for providers of generative models to meet legal requirements to mark synthetic content.
A watermark embeds a hidden signal into generated content that a detector holding a secret key can later recover.
Its security is evaluated primarily against adversaries who remove or forge this signal in individual outputs.
We argue that watermarking should also be evaluated and governed as a \emph{monitoring primitive}, because the same signal can link outputs to the entity that generated them.
We show that (i) internal observers with detector access attribute outputs to individual entities even under zero-bit watermarks, which encode no identity, when each entity holds its own key, and (ii) external observers without keys or detectors identify the source of unseen text and image outputs in the configurations we test.
Both depend on deployment choices that evaluations do not report, namely how fine-grained and persistent the bindings between keys and entities are and who may query the detector.
We call for evaluations to report attribution and identification accuracy alongside robustness, and for governance to limit these bindings and detector access.
\end{preprintabstract}

\section{Introduction}

Regulators now require providers of generative models to mark synthetic content in a machine-readable form \citep{eu_2024, CA_SB942_2024}, and major providers have committed to watermarking their outputs \citep{bartz_hu_2023}.
A watermark embeds a hidden signal into generated content that a detector holding a secret key can later recover \citep{zhao2024sok}.
The signal serves provenance \citep{Pang2024NoFL, zhou2024bileve}, attribution of outputs to their source \citep{aaronson2023watermarking, kirchenbauer2023watermark, liu2024an, hou2023semstamp, synthid} and safety monitoring \citep{aremu2026robust}.
When a provider assigns each account its own key \citep{aremu2025mitigating} or encodes an account identifier in the watermark \citep{wang2024towards}, the same detector also reveals which account generated an output, as \Cref{fig:concept} illustrates.

\begin{figure}[t]
    \centering
\begin{tikzpicture}[
    x=1cm, y=1cm, font=\footnotesize,
    >={Stealth[length=4pt]},
    doc/.style={draw=black!60, fill=white, rounded corners=1pt, minimum width=1.55cm, minimum height=0.78cm, align=center, inner sep=1.5pt},
    box/.style={draw=black!70, rounded corners=2pt, align=center, inner sep=3pt},
    tag/.style={rounded corners=2pt, inner sep=1.6pt, font=\scriptsize},
    lane/.style={rounded corners=3pt, line width=0.6pt},
    lev/.style={circle, draw=leverc, fill=leverc!12, text=leverc, inner sep=0.8pt, font=\scriptsize\bfseries},
    levtxt/.style={text=leverc, font=\scriptsize, align=center},
]
\node[box, minimum width=1.55cm, minimum height=0.95cm] (ent) at (0.75,0) {Entity $e$\\[-1pt]{\scriptsize key $k_e$}};
\node[box, fill=black!4, minimum width=1.9cm, minimum height=0.95cm] (model) at (3.0,0) {Watermarked\\[-1pt]model $W_{k_e}$};
\draw[->] (ent) -- (model);
\node[lev] (l1) at (0.75,-0.85) {1};
\node[levtxt, anchor=west] at (l1.east) {granularity};

\foreach \i/\x/\venue in {1/5.3/service, 2/7.1/forum, 3/8.9/news site, 4/10.7/anonymous post} {
    \node[doc] (x\i) at (\x,0) {$x_\i$\\[-1pt]{\scriptsize \venue}};
}
\draw[->] (model) -- (x1);
\draw[->, black!45] (4.45,-0.6) -- (11.55,-0.6) node[right, black!60, font=\scriptsize] {time};
\node[lev] (l2) at (8.0,-0.6) {2};
\node[levtxt, anchor=west, fill=white, inner sep=1pt] at (l2.east) {persistence};

\draw[lane, draw=internalc, fill=internalc!5] (4.35,0.72) rectangle (11.55,1.62);
\node[text=internalc, align=center, font=\scriptsize\bfseries] at (2.25,1.17) {Internal observer};
\node[text=internalc, align=center, font=\scriptsize] at (2.25,0.86) {detectors + key mapping};
\node[lev] (l3) at (0.45,1.17) {3};
\node[levtxt, anchor=north] at (0.45,1.0) {access};
\foreach \i in {1,...,4} {
    \draw[->, internalc] (x\i.north) -- ++(0,0.45);
    \node[tag, fill=internalc!15, text=internalc] at ($(x\i.north)+(0,0.68)$) {attributed to $e$};
}
\draw[->, internalc, line width=0.8pt] (11.55,1.17) -- (11.95,1.17);
\node[text=internalc, font=\scriptsize, align=left, anchor=west] at (11.95,1.17) {observed\\[-1pt]activity of $e$};

\draw[decorate, decoration={brace, mirror, amplitude=3pt}, externalc] (4.55,-0.82) -- (7.85,-0.82)
    node[midway, below=2pt, font=\scriptsize, text=externalc] (lab) {labeled outputs (known account)};
\draw[lane, draw=externalc, fill=externalc!5] (4.35,-2.55) rectangle (11.55,-1.45);
\node[text=externalc, align=center, font=\scriptsize\bfseries] at (2.25,-1.85) {External observer};
\node[text=externalc, align=center, font=\scriptsize] at (2.25,-2.16) {no keys or detectors};
\node[box, draw=externalc, fill=white, text=externalc, font=\scriptsize] (clf) at (8.9,-1.92) {classifier $f$};
\draw[->, externalc] (6.2,-1.3) -- (6.2,-1.92) -- (clf.west);
\draw[->, externalc] (x4.south) -- (10.7,-1.92) -- (clf.east);
\node[tag, fill=externalc!15, text=externalc, anchor=north] at (8.9,-2.2) {$x_4$ identified as $e$};
\end{tikzpicture}
    \caption{\small \textbf{Repeated attribution turns a watermark into a monitoring primitive.} An entity $e$ generates outputs under its key $k_e$, and the outputs appear beyond the provider's service. An internal observer with detectors and the key mapping attributes every output to $e$, which yields a record of $e$'s observed activity. An external observer without keys learns from outputs published under a known account and identifies the source of a later anonymous output. Orange numbers mark where governance acts, namely \lever{1} the granularity of key bindings, \lever{2} their persistence and \lever{3} access to detectors and the key mapping.}
    \label{fig:concept}
\end{figure}

Consider a provider that grants an authority detector access and the mapping from keys to accounts for one investigation, as in the scenarios of \Cref{fig:scenarios}.
If this access and the keys outlive the investigation, the authority can attribute every later output it encounters to the same accounts, including outputs published outside the service.
An outsider without detector access can instead collect outputs published under known accounts and learn to recognize each account's watermark in later anonymous outputs.
Account logs offer neither capability, because they stay with the provider, whereas the watermark remains in the output after it leaves the service.
How far either actor can follow an account depends on the granularity and persistence of key bindings, on access to the detector and on how learnable the watermark is.

Watermark security is evaluated primarily against adversaries who evade detection \citep{diaa2024optimizing, lukas2024leveraging, pang2024attacking, wu2024bypassing, krishna2023paraphrasing} or forge watermarks to cause false positives \citep{jovanovic2024watermarkstealing, gloaguen2024discovering, aremu2025mitigating, muller2025black}.
These evaluations measure whether the signal in an individual output survives paraphrasing, rewriting or other transformations \citep{kirchenbauer2023reliability, pan2024markllm, Piet2023MarkMW, zhao2024provable, christ2024provably}.
However, they do not measure what an observer learns by using the signal, rather than removing or forging it, across many outputs of the same entity.

\textbf{Position.} We argue that watermarking should be evaluated and governed as a \emph{monitoring primitive}, accounting for what observers can infer about the entities, such as accounts or organizations, behind its outputs.
We distinguish \emph{internal observers}, who can query detectors and know which key belongs to which entity, from \emph{external observers}, who have neither and learn entity-specific signals from outputs with known sources.
\emph{Multi-bit} watermarks encode identifiers that make internal attribution immediate \citep{wang2024towards}, so we focus on \emph{zero-bit} watermarks, which encode only their presence.
Attribution can serve legitimate audits and investigations, so the question for governance is who may repeat it, for which purposes and for how long.

We evaluate twelve zero-bit methods, nine for text on \texttt{Qwen2.5-14B} and three for images on \texttt{Stable Diffusion v2.1}, with up to 16 entities that each hold their own key.
For every method, an internal observer attributes at least 96\% of outputs to the correct entity at a 1\% false-positive rate per key.
An external observer without keys or detectors identifies the source of held-out text outputs watermarked with KGW \citep{kirchenbauer2023watermark} among 16 entities with 73.0\% top-1 accuracy after training on up to 4000 labeled outputs per entity, against 6.25\% for random guessing.
Images watermarked with Tree-Ring \citep{wen2023tree} reach 91.0\% with the same budget, and both rates fall to near chance without a watermark or under a shared key, so identification requires key-dependent structure, which we establish for one text and one image method.

\textbf{Contributions.}
We
(i) propose an observer-based threat model that separates attribution through detector access from identification without keys or detectors (\Cref{sec:threat}),
(ii) demonstrate both pathways, with no-watermark and shared-key controls that isolate the contribution of the watermark (\Cref{sec:experiments}), and
(iii) propose that evaluations report attribution and identification accuracy as functions of the number of entities and of the labeled outputs per entity, and identify the granularity and persistence of key bindings and access to detectors as the levers for governance (\Cref{sec:discussion}).

\section{Background}

\textbf{Watermarking.}
Watermarking embeds a detectable signal into generated content to enable downstream verification \citep{zhao2024sok}.
Watermarks may be \emph{zero-bit}, indicating only presence or absence, or \emph{multi-bit}, encoding additional information such as identifiers \citep{zhao2024sok, wang2024towards}.
A key design goal is robustness, i.e., the watermark should remain detectable under transformations \citep{kirchenbauer2023reliability, christ2024provably}.

\textbf{Threat Models for Watermarking.}
Prior work primarily studies watermarking under adversaries who aim to disrupt or exploit the watermark signal.
Common threats include \emph{evasion}, where outputs are modified to remove the watermark \citep{krishna2023paraphrasing, diaa2024optimizing, pang2024attacking},
\emph{forgery}, where non-watermarked content is attributed to a watermarked source \citep{aremu2025mitigating, jovanovic2024watermarkstealing, gloaguen2024discovering, muller2025black},
and \emph{secret extraction}, where the watermarking key or decision boundary is inferred \citep{gu2024on}.
These objectives can also involve learning from many outputs, as watermark stealing uses model queries to approximate watermark rules for spoofing or removal \citep{jovanovic2024watermarkstealing}.
Our work differs in its objective rather than in aggregation, since detecting a watermark, recovering its rules and identifying the entity that generated an output are distinct tasks, and success at the first two does not by itself establish the third.

\textbf{Watermarking for Monitoring.}
Recent work uses watermarking for safety monitoring, embedding signals into model behavior to detect policy-violating outputs \citep{aremu2026robust}.
Text forensics studies detection, attribution and characterization of generated text \citep{kumarage2024survey}, including stylometric detection across timelines \citep{kumarage2023stylometric} and neural authorship attribution \citep{kumarage2023neural}.
Our work instead studies the monitoring capability that deploying a watermark creates, and our no-watermark and shared-key controls separate it from authorship signals in the content itself.

\textbf{Identity Encoding and Privacy.}
Multi-bit schemes explicitly encode recoverable information \citep{wang2024towards, yoo2024advancing}.
Prior work also warns that embedded user identifiers and detector access can link anonymous publications to an identity \citep{zhao2024sok}.
We build on these concerns rather than claim the first observation that watermarking can enable identification, and connect the governance of attribution access with design-dependent external identification.

\begin{figure*}[t]
\centering
\begin{tcolorbox}[
    enhanced,
    colback=white,
    colframe=themeIndigo, 
    boxrule=0.8pt,
    arc=2mm,
    left=2mm,
    right=2mm,
    top=1.5mm,
    bottom=1.5mm,
    title=\textbf{Example of Monitoring Scenarios Enabled by Watermarking},
    fonttitle=\bfseries\color{white},
    coltitle=white,
    colbacktitle=themeIndigo, 
    width=\textwidth
]
\renewcommand{\arraystretch}{1.25}
\begin{tabularx}{\textwidth}{>{\raggedright\arraybackslash}X >{\raggedright\arraybackslash}X}
\begin{tcolorbox}[
    enhanced,
    colback=white,
    colframe=themeIndigo, 
    boxrule=0.6pt,
    arc=1.5mm,
    title=\textbf{Provider-side attribution},
    fonttitle=\bfseries\color{white},
    colbacktitle=themeIndigo, 
    height=3.5cm
]
\small A provider assigns persistent keys to accounts and holds the key mapping.
With detector access, it can attribute observed outputs for auditing or abuse investigation, including outputs encountered outside the service.
\end{tcolorbox}
&
\begin{tcolorbox}[
    enhanced,
    colback=white,
    colframe=themeIndigo,
    boxrule=0.6pt,
    arc=1.5mm,
    title=\textbf{External identification},
    fonttitle=\bfseries\color{white},
    colbacktitle=themeIndigo,
    height=3.5cm
]
\small A third party learns from labeled outputs without keys or detectors.
It may identify the source of unseen outputs when key bindings persist and the watermark induces learnable structure.
\end{tcolorbox}
\\
\begin{tcolorbox}[
    enhanced,
    colback=white,
    colframe=themeIndigo,
    boxrule=0.6pt,
    arc=1.5mm,
    title=\textbf{Linking anonymous publications},
    fonttitle=\bfseries\color{white},
    colbacktitle=themeIndigo,
    height=3.5cm
]
\small An external observer could use outputs reliably associated with a known account to identify later anonymous publications.
This scenario requires suitable reference outputs and persistent, learnable watermark signals; it need not reveal a real-world identity.
\end{tcolorbox}
&
\begin{tcolorbox}[
    enhanced,
    colback=white,
    colframe=themeIndigo,
    boxrule=0.6pt,
    arc=1.5mm,
    title=\textbf{Surveillance by an authority},
    fonttitle=\bfseries\color{white},
    colbacktitle=themeIndigo,
    height=3.5cm
]
\small An authority granted detector access and the key mapping is an internal observer.
If access and bindings persist, repeated attribution can support surveillance of observed activity without direct account access.
\end{tcolorbox}
\end{tabularx}
\end{tcolorbox}
\caption{\small \textbf{Hypothetical scenarios in which watermarks support monitoring.} Provider-side attribution and surveillance by an authority require detector access (internal observers), whereas the other two scenarios learn from labeled outputs (external observers).}
\label{fig:scenarios}
\end{figure*}

\section{Threat Model} \label{sec:threat}

We consider a provider that serves a watermarked generative model to many entities, such as user accounts or organizations, and an observer who uses the watermark in their outputs to infer which entity generated them.
The observer is passive, as it neither controls generation nor modifies outputs, and \Cref{fig:scenarios} illustrates four hypothetical scenarios.

\textbf{Setting.}
Let $\mathcal{M}$ denote a generative model that maps a prompt $p \in \mathcal{P}$ to an output $x \in \mathcal{X}$, where $x \sim \mathcal{M}(\cdot \mid p)$.
A watermarking scheme embeds a signal under a secret key $k \in \mathcal{K}$ with an embedding function $W_k$, so that $x \sim W_k(\mathcal{M}(\cdot \mid p))$, and a detector computes a score $\mathcal{D}_k(x)$ that marks $x$ as watermarked when it exceeds a threshold $\tau$.
Entities $\mathcal{E} = \{e_1, \dots, e_n\}$ generate outputs over time.
Under \emph{multi-key deployment} \citep{aremu2025mitigating}, the provider assigns each entity $e$ its own key $k_e$, which we call the entity's \emph{key binding}, and the record of all bindings is the \emph{key mapping}.
A \emph{multi-bit} scheme instead embeds a message $m \in \{0,1\}^b$ that can identify the entity \citep{wang2024towards}.

\textbf{Internal Observer.}
\textit{(Access)} An internal observer can evaluate the detectors, through key possession or a detector interface, and holds the key mapping.
We classify observers by this access rather than by affiliation, so a provider, platform or authority granted it is an internal observer.
\textit{(Persistence)} While key bindings persist, the same access allows repeated attribution of an entity's outputs, provided detection reliably distinguishes keys.

\textbf{External Observer.}
\textit{(Access)} An external observer has neither keys, detector access nor the key mapping.
\textit{(Data)} It instead collects \emph{labeled outputs}, i.e., outputs with known sources, such as outputs published under known account handles or organizational bylines, and uses them to identify the source of later anonymous outputs.
We assume reliable labels and the same key bindings for labeled and anonymous outputs, as an observer cannot learn a binding that changes between the two.
A publication account need not identify a verified person or the generating account.
Our experiments assume suitable labels and do not establish that thousands of such outputs are available for a typical user.
Labeled and anonymous outputs are disjoint, since an observer that has already seen an output with its label identifies it trivially.
\textit{(Inference)} The observer trains a classifier on labeled outputs and predicts the source of each anonymous output individually, without joint inference over several anonymous outputs.

\textbf{Observer's Goals.}
Both observers pursue \emph{identification}, which assigns an output $x$ to the entity $e(x)$ that generated it and which we call \emph{attribution} when an internal observer uses detectors.
Repeated identification serves two further goals, (i) \emph{linkability}, which decides whether $e(x_i) = e(x_j)$ for two outputs, and (ii) \emph{monitoring}, which tracks an entity's activity within the observed outputs rather than its total model usage.

\section{Watermarking as a Monitoring Primitive} \label{method}

\begin{figure}[t]
    \centering
    \includegraphics[width=\linewidth]{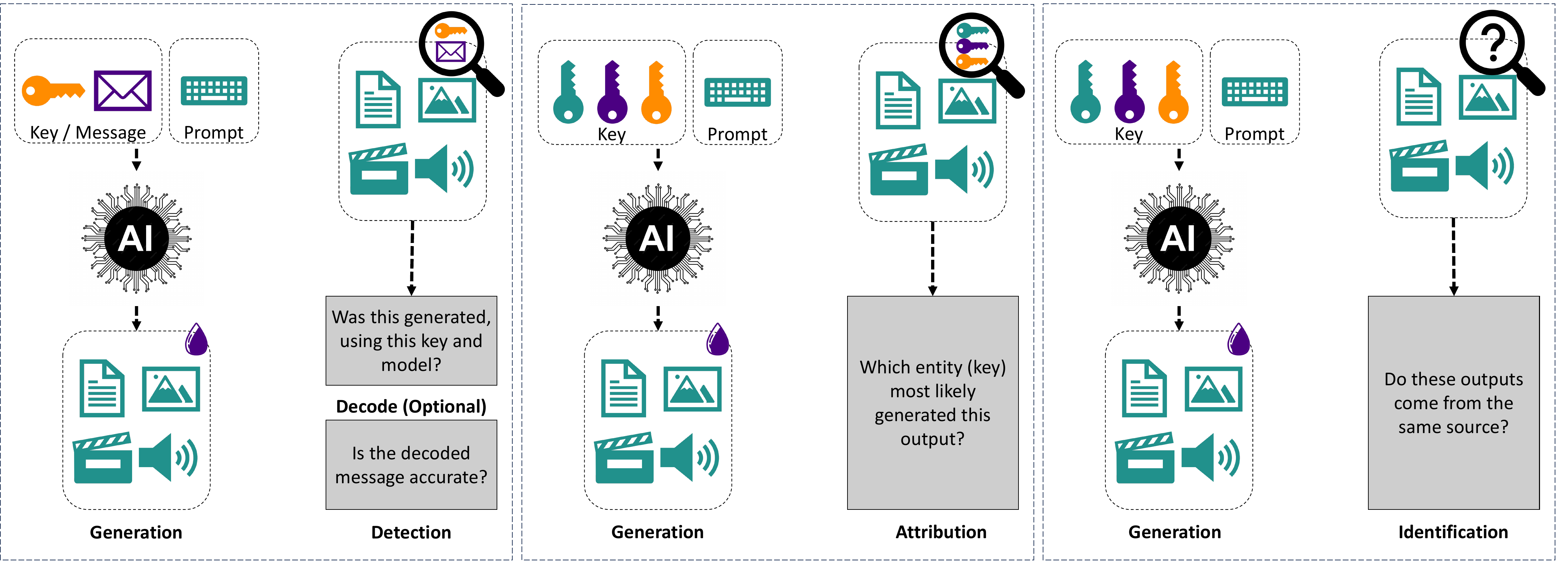}
    \caption{\small \textbf{The signal that verifies provenance can also attribute outputs to entities.} \textbf{Left:} \textit{Standard watermarking}, where a detector determines whether an output is watermarked and optionally decodes an embedded message. \textbf{Middle:} \textit{Internal observer}, who has access to watermark keys and performs attribution by identifying which entity generated an output. \textbf{Right:} \textit{External observer}, who does not have access to keys and instead learns to identify which entity generated an output from publicly observed data by exploiting watermark-induced statistical patterns.}
    \label{fig:teaser}
\end{figure}

Each observer of \Cref{sec:threat} turns the watermark into identification without a new attribution algorithm, the internal observer by repeating detection and the external observer by learning the structure each key leaves in outputs, as \Cref{fig:teaser} shows.

\textbf{From Detection to Attribution.}
Watermarks are usually evaluated per output, as the detector tests whether one output $x$ carries the watermark of one key $k$.
Multi-bit watermarks make attribution immediate, since the detector decodes a message $\hat m$ that the key mapping relates to an entity \citep{wang2024towards}.
Zero-bit watermarks encode no message, but under multi-key deployment each key $k_e$ induces its own output distribution $W_{k_e}(\mathcal{M}(\cdot \mid p))$.
An internal observer therefore evaluates every detector and selects
\begin{equation}
    \hat{e}(x) = \arg\max_{e \in \mathcal{E}} \mathcal{D}_{k_e}(x),
\end{equation}
accepting the attribution only if the winning score also passes the calibrated threshold of its key, and abstaining otherwise.
Repeating this attribution while key bindings persist monitors the entity's observed outputs, and two outputs are linked when both are attributed to the same entity.

\textbf{From Labeled Outputs to Identification.}
An external observer has no detectors, so it relies on the statistical structure that each key leaves in outputs.
Given labeled outputs $\{(x_i, e_i)\}_{i=1}^N$ and observable features $\phi(x)$, such as lexical or embedding-based features, it trains a classifier $f$ and predicts
\begin{equation}
    \hat e(x) = f(\phi(x)).
\end{equation}
This identification succeeds only when key-dependent structure persists across labeled and anonymous outputs and the classifier can learn it, which \Cref{sec:experiments:external} tests for KGW and Tree-Ring.

\section{Experiments}
\label{sec:experiments}

Our experiments address three questions, namely (1) whether an internal observer can attribute outputs to entities under zero-bit watermarks, (2) whether an external observer without keys or detectors can identify their source, and (3) whether the watermark, rather than content or prompts, causes this identification.

\textbf{Watermarking Methods.}
We evaluate twelve \emph{zero-bit} watermarking methods.
For text, we instantiate nine methods \citep{kirchenbauer2023watermark, christ2024undetectable, zhao2024provable, lu2024entropy, wang2025morphmark, hou2023semstamp, synthid, gu2025invisible, lee2024wrote} from \textbf{MarkLLM} \citep{pan2024markllm}.
For images, we instantiate three methods \citep{wen2023tree, yang2024gaussian, arabi2024hidden} from \textbf{MarkDiffusion} \citep{pan2025markdiffusion}.

\textbf{Implementation Configuration.}
We primarily use the default watermark configurations provided by MarkLLM and MarkDiffusion, with the generation models specified below and entity-specific keys in the multi-key setting.
For KGW, these defaults specify green-list fraction $\gamma=0.5$, logit bias $\delta=2.0$, context width $h=1$, and the \texttt{time} function and \texttt{left} window schemes.\footnote{MarkLLM's \texttt{config/KGW.json}, revision \href{https://github.com/THU-BPM/MarkLLM/blob/32048437234101f210d3d6e6ef488834b86fbbcd/config/KGW.json}{\texttt{3204843}}. This identifies the configuration source, not the experimental software checkout.}
The entity-specific keys and calibrated detection thresholds replace the corresponding default settings; the toolkit's fixed KGW $z$-threshold is not the threshold used for the reported 1\% FPR operating point.
For Tree-Ring, entities receive distinct combinations of watermark pattern, radius, mask shape, and channel.
Here, an entity's ``key'' includes this configuration, rather than only a random seed; these results therefore concern configuration-level distinctions, not seed-only keys under a fixed watermark configuration.


\textbf{Models.}
We use \texttt{Qwen2.5-14B} \citep{qwen2.5} for text and \texttt{Stable Diffusion v2.1} \citep{rombach2022high} for image generation.
For each modality, we study watermarking under standard deployment and under \emph{multi-key deployment}, where each entity holds a distinct key.


\textbf{Datasets.}
We use shared prompt pools across entities, so that prompt differences cannot explain attribution or identification.
For text, we use C4 \citep{2019t5}, a cleaned subset of Common Crawl spanning multiple content categories.
For images, we use the Stable Diffusion Prompt dataset\footnote{\url{https://huggingface.co/datasets/Gustavosta/Stable-Diffusion-Prompts}}.
All entities generate outputs from the same prompts under different keys.
Training and evaluation outputs come from disjoint prompts, so no prompt or output appears in both splits.
%


\textbf{Metrics.}
For internal observers, we report the \textbf{correct-and-detected attribution rate}, denoted TPR@1\%FPR in \Cref{fig:internal}.
Concretely, for each key $k$, we calibrate a detection threshold $\tau_k$ to achieve a 1\% false positive rate on non-matching samples (i.e., outputs generated under other keys).
Given an output $x$, we select the entity $\hat e(x)$ whose key yields the highest detector score, and accept the attribution only if that score exceeds $\tau_{k_{\hat e(x)}}$; otherwise, we abstain.
The reported rate is the fraction of all evaluated outputs for which the winning key is correct and its threshold is passed; both incorrect predictions and abstentions count as failures.
The 1\% false positive rate is controlled per key and does not directly correspond to a global false positive rate under selection among many keys, as correlations between detector scores may affect attribution at larger scales.
For external observers, we report standard \textbf{top-1} and \textbf{top-3 classification accuracy}, where uniform random guessing corresponds to $1/n$ and $\min(3/n,1)$, respectively, for $n$ entities.
External observer evaluation is performed on a held-out set of 100 samples per entity.


\subsection{Internal Observer: Attribution under Zero-Bit Multi-Key Watermarking}
\label{sec:experiments:internal}

\begin{figure}[ht!]
\centering
\includegraphics[width=\linewidth]{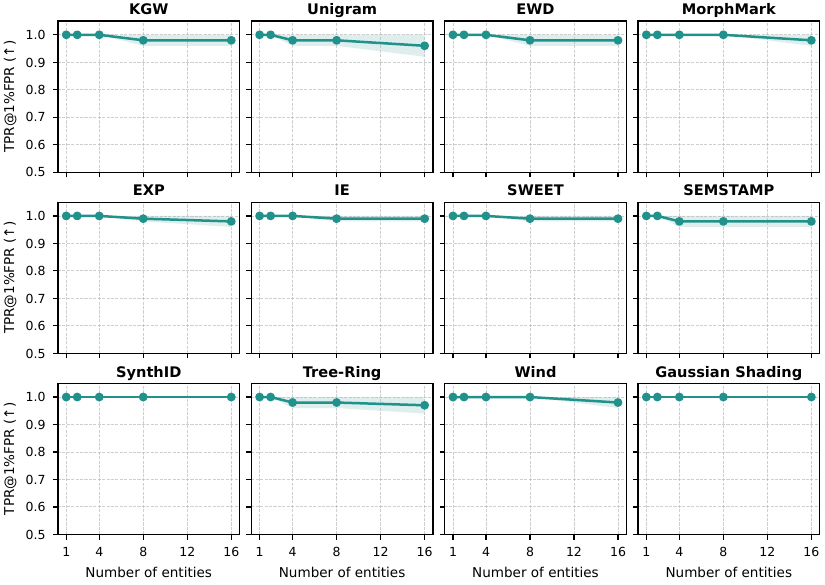}
\caption{\small \textbf{Internal attribution stays at or above 96\% for all twelve methods with up to 16 entities.} Success requires the correct key to score highest and pass its calibrated threshold. TPR@$1\%$FPR uses per-key calibration, not a global 1\% false-attribution guarantee. Shaded bands show one standard deviation across runs.}
\label{fig:internal}
\end{figure}

An internal observer holds a detector for each entity's key and attributes an output to the entity whose detector scores highest, provided that score passes the key's threshold.
\Cref{fig:internal} plots the correct-and-detected attribution rate against the number of candidate entities, from 1 to 16, with 100 evaluation outputs per entity.
\textbf{Every method attributes at least 96\% of outputs correctly at 16 entities}, and the lowest rates occur for Unigram and Tree-Ring.
Zero-bit watermarks therefore support attribution without encoding an identity once each entity holds its own key.


\subsection{External Observer: Identification from Labeled Outputs}
\label{sec:experiments:external}

An external observer has neither keys nor detectors and instead trains a classifier on outputs with known sources to predict the source of held-out outputs.
We consider $n \in \{2,4,8,16\}$ entities with distinct keys under KGW for text and Tree-Ring for images.
External coverage is narrower than the twelve-method internal evaluation, because each setting requires labeled generation and classifier training, so the two methods are text and image feasibility cases within our computational budget, not representative coverage of all watermark designs.
The observer trains BERT-Base \citep{DBLP:journals/corr/abs-1810-04805} for text and CLIP-RN50 \citep{radford2021learning} for images on 100 to 4000 labeled outputs per entity (batch size 16, 10 epochs, AdamW/Adam, learning rate $3 \times 10^{-5}$).

\begin{figure}[ht!]
    \centering
    \includegraphics[width=1\linewidth]{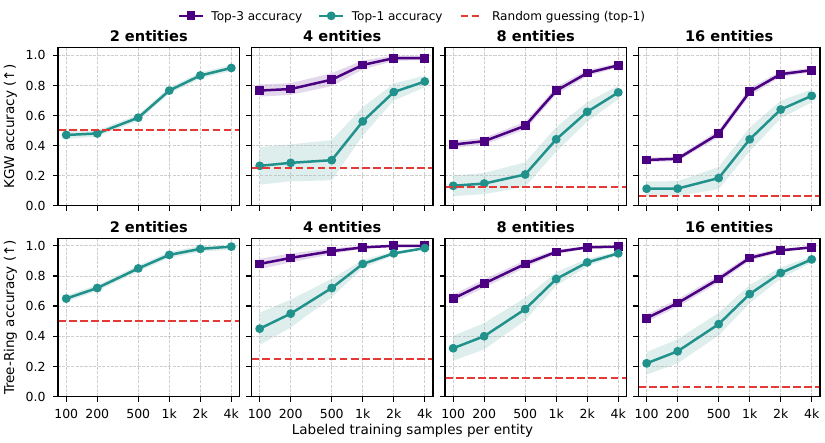}
    \caption{\small \textbf{External identification improves with the number of labeled outputs per entity for both KGW text and Tree-Ring images.} Top-1 and top-3 accuracy are plotted against labeled training samples per entity, on a log scale, for $n \in \{2,4,8,16\}$; their uniform-guessing baselines are $1/n$ and $\min(3/n,1)$. Dashed lines show the top-1 baseline, and shaded bands one standard deviation across runs. The sample axis is training-set size, not elapsed time or joint inference over several anonymous outputs.}
    \label{fig:external}
\end{figure}

\Cref{fig:external} plots top-1 and top-3 accuracy against the labeled outputs per entity, with dashed lines marking top-1 random guessing.
\textbf{Identification starts near chance and improves as labeled outputs grow.}
With 16 KGW entities, top-1 accuracy rises from 11.3\% to 73.0\%, against 6.25\% for random guessing, and top-3 accuracy reaches 90.0\%.
Tree-Ring reaches higher accuracy with fewer labeled outputs, with 91.0\% top-1 accuracy for 16 entities at 4000 outputs per entity.

\textbf{Targeted Identification.}
An observer may instead ask whether an output comes from one known target.
We train a binary KGW classifier for one randomly selected target, pooling the other 15 entities into a background class, and test it on 100 target and 1,500 background outputs.

\begin{wraptable}{r}{0.46\textwidth}
\vspace{-0.75em}
    \centering
    \small
    \caption{\textbf{A targeted observer finds every target output at a 3.2\% false positive rate.} Rates treat each class as positive in turn. Overall accuracy is 97.0\%.}
    \label{tab:targeted}
    \setlength{\tabcolsep}{12pt}
    \begin{tabular}{@{}lrrr@{}}
        \toprule
        Class & Samples & TPR & FPR \\
        \midrule
        Target & 100 & 100\% & 3.2\% \\
        Pooled & 1,500 & 96.8\% & 0\% \\
        \bottomrule
    \end{tabular}
\vspace{-0.75em}
\end{wraptable}
The classifier identifies all 100 target outputs and assigns 48 background outputs to the target, which gives 97.0\% overall accuracy, as \Cref{tab:targeted} shows.
Always predicting the background class would reach 93.75\% accuracy with zero target recall.
This result covers one target against 15 background entities, so it does not establish reduced training requirements or effectiveness against larger, more diverse populations.
The nonzero false positive rate also limits interpretation when target outputs are rare.

\begin{figure}[t]
    \centering
    \includegraphics[width=1\linewidth]{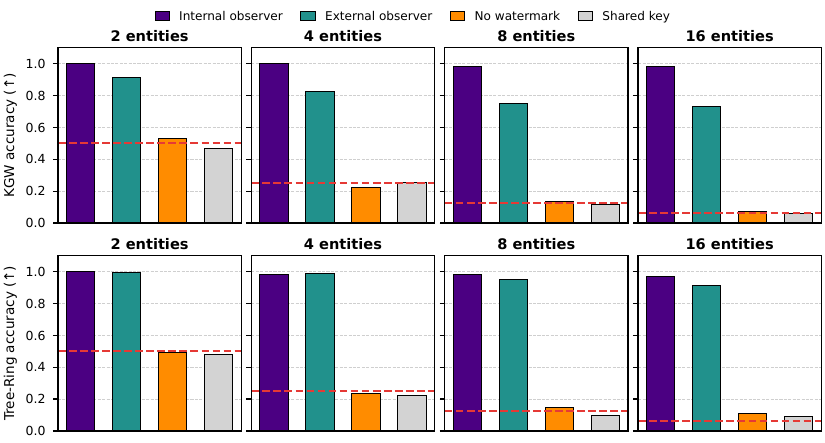}
    \caption{\small \textbf{Without a watermark or with a shared key, identification falls to near chance.} We compare four settings: internal observer (with key access), external observer (learned classifier), no watermark, and shared-key deployment. Results are shown for both text (KGW) and image (Tree-Ring) watermarking across $n \in \{2,4,8,16\}$ entities, with random guessing indicated by the dashed baseline ($1/n$). Internal attribution stays near 100\% under multi-key deployment, and external identification is lower. Both controls stay near random guessing, indicating that identification comes from key-dependent structure, not from content or prompts.}
    \label{fig:control}
\end{figure}

\textbf{Controls.}
Two controls test whether the watermark causes external identification, as \Cref{fig:control} shows.
In the \emph{no-watermark} control, outputs carry no watermark but keep their entity labels, and identification accuracy stays near random guessing, with small deviations at small $n$ due to finite samples and residual classifier bias.
In the \emph{shared-key} control, all entities use the same key, and accuracy again stays near chance.
Under the matched model, prompts and generation settings, sharing a key thus removes the watermark-based distinction between entities, although it does not establish anonymity where other identifying signals are present.

\section{Discussion}
\label{sec:discussion}

\textbf{Two Pathways from Attribution to Monitoring.}
Attribution access and external identification raise two distinct governance questions, namely how those with access exercise it and what outsiders can infer without it.
For an internal observer, one attribution can serve an audit or an investigation, whereas repeated attribution under persistent per-entity keys follows an entity's observed outputs, even in zero-bit schemes.
The concern is not each attribution query but that access granted for one purpose also enables sustained observation, a capability we establish rather than observed misuse.

\textbf{Governing Granularity, Persistence and Access.}
The threat model of \Cref{sec:threat} exposes three deployment choices that governance can set.
\emph{First}, shared or group-level keys reduce what the watermark reveals to the group, at the cost of entity-specific attribution, although other signals may still identify the entity.
\emph{Second}, key rotation limits linkage across rotations only when the observer cannot access the mappings between old and new keys.
\emph{Third}, restricting access to detectors, decoders and key mappings limits who can attribute and for which purposes, but not what those who retain access can do.
We have not tested these choices experimentally, and none of them requires abandoning provenance or auditing.

\textbf{From Provenance Compliance to Monitoring Governance.}
Existing transparency frameworks already address privacy in provenance infrastructure, as the EU's transparency Code sets purpose, retention and access safeguards for detection services \citep[Sub-measures 2.1.2--2.1.3]{eu_transparency_code_2026} and California's AI Transparency Act restricts the disclosure and retention of personal provenance data \citep{CA_BPC22757_2}.
We argue that such safeguards should also cover attribution access and the entity-specific structure that outsiders can learn.
Governance must specify who may authorize repeated attribution, for which purposes and under what oversight, whether providers and authorities may retain key mappings or accumulate attribution results across contexts, and how affected users can contest such practices.
Provenance compliance should therefore come with an explicit assessment of entity linkability and enforceable limits on its use, since identifying that content is AI-generated differs from monitoring the entity that generated it.

\textbf{Watermark Design and External Exposure.}
Restricting detector access does not stop outsiders from learning entity-specific signals, so design selection should weigh external linkability alongside provenance performance.
We hypothesize that distribution-preserving and undetectable constructions \citep{christ2024pseudorandom, christ2024undetectable, gunn2024undetectable} weaken external identification \citep{zhao2024sok}, as they limit the key-dependent structure an observer can learn.
We have not tested this hypothesis, and preserving the distribution of a single output does not by itself guarantee protection across repeated observations.
Black-box watermark inference \citep{gloaguen2025blackbox} and watermark stealing \citep{jovanovic2024watermarkstealing, pang2024attacking} do not show entity identification under these constructions, and our experiments do not show that their guarantees fail.

\textbf{Towards Weaker Supervision and Larger Populations.}
Our external observer trains on outputs with known sources and assumes the same key bindings at training and test time.
A targeted observer, which asks whether an output belongs to one known entity, reaches 100\% recall at a 3.2\% false-positive rate against 15 pooled background entities, as \Cref{tab:targeted} shows, whereas larger or more diverse backgrounds remain untested.
Semi-supervised or clustering observers could relax the labeling assumption, which we leave to future work, so our results do not establish monitoring at population scale or without labeled data.
Neither our experiments nor our scenarios reconstruct complete activity histories or observe actual surveillance.

\textbf{Towards Identification under Transformations.}
We do not test external identification under decoding changes, under post-processing such as paraphrasing and summarization, which can affect detectability \citep{pan2024markllm, Piet2023MarkMW}, or under image editing.
Our KGW configuration uses a context width of $h=1$, and without a sweep over context width or watermark strength, other settings remain open.
Whether more labeled outputs compensate for weaker signals, and whether joint inference over several anonymous outputs strengthens identification, also remain untested.
Internal attribution after such transformations likewise depends on detection remaining reliable, not on detector access alone.

\textbf{Recommendations for Evaluation.}
Watermark evaluations should measure monitoring risk under both observer models, alongside robustness and detectability.
At a minimum, they should report attribution and identification accuracy as functions of the number of entities and of the outputs available per entity, separating labeled training outputs from repeated anonymous observations.
Deployments should also disclose the granularity and persistence of their key bindings and who can access detectors and key mappings.

\section{Alternative Views}
\label{sec:alternative}

\textbf{Providers already log every request.}
A provider that logs which account issued each request can attribute its own outputs without a watermark, which suggests that watermarking adds no monitoring capability.
However, logs stay with the provider and cover only requests to its service, whereas the watermark remains in the output wherever it is published.
Watermark attribution therefore reaches outputs found outside the service and any party granted detector access, and external identification needs no access to the provider at all.

\textbf{Attribution is the purpose of watermarking, not a risk.}
Provenance checks and abuse investigations need attribution, and restricting it would weaken the accountability that motivates watermarking.
We agree that an authorized attribution for a stated purpose is the intended use.
Our position concerns its repetition, since the access that serves one investigation can follow an entity across contexts unless governance limits its purpose, retention and scope.

\textbf{Undetectable watermarks prevent identification.}
Watermarks whose outputs no efficient observer can distinguish from unwatermarked outputs without the key \citep{christ2024undetectable} may prevent external identification, which we have not tested.
We agree that such designs would remove the external pathway, which is why design selection should weigh linkability.
They leave internal attribution to anyone with reliable detection, persistent bindings and detector access, so its governance applies to every design.

\section{Conclusion}

We argue that watermarking should be evaluated and governed as a monitoring primitive as well as a provenance mechanism.
Across twelve zero-bit methods and up to 16 entities, internal attribution is correct for at least 96\% of outputs at a 1\% false-positive rate per key.
Without keys or detectors, an external observer identifies the source among 16 entities with 73.0\% top-1 accuracy for KGW text and 91.0\% for Tree-Ring images, whereas outputs without a watermark or under a shared key leave it near chance, which we establish for one text and one image method.
We therefore call for watermark evaluations to report attribution and identification accuracy under both observer models, and for governance to limit key bindings and detector access, so that provenance and auditing keep their legitimate uses.

\section*{Ethical Considerations}

This work examines internal attribution access and design-dependent external identification.
Both can support monitoring when key bindings persist and inference is reliable.
We neither advocate unjustified surveillance nor establish actual misuse.
We seek to inform governance through transparency about entity-identifying capabilities, limits on secondary use, and opportunities for affected users to contest consequences.

\bibliographystyle{abbrvnat}
\small{\bibliography{ref}}

\end{document}